\documentclass[prl, aps, twocolumn, preprintnumbers,showpacs, superscriptaddress]{revtex4}

\usepackage{latexsym}
\usepackage{mathrsfs}
\usepackage{amsmath}
\usepackage{amssymb}
\usepackage{amsfonts}
\usepackage{ifthen}

\newcommand{\proj}[1]{\ket{#1}\bra{#1}}

\newcommand{\be}{\begin{equation}}
\newcommand{\ee}{\end{equation}}
\newcommand{\bea}{\begin{eqnarray}}
\newcommand{\eea}{\end{eqnarray}}

\newcommand{\hilbert}{\mathcal{H}}
\newcommand{\cY}{{\cal Y}}      
\newcommand{\cU}{{\cal U}}      

\newcommand{\cE}{{\cal E}}

\newcommand{\I}{I_{acc}}        
\newcommand{\ens}{{\cal E}}
\newcommand{\nc}{\newcommand}
\nc{\RR}{{{\mathbb R}}}
\nc{\CC}{{{\mathbb C}}}
\nc{\FF}{{{\mathbb F}}}
\nc{\NN}{{{\mathbb N}}}
\nc{\ZZ}{{{\mathbb Z}}}
\nc{\PP}{{{\mathbb P}}}
\nc{\QQ}{{{\mathbb Q}}}
\nc{\UU}{{{\mathbb U}}}
\nc{\EE}{{{\mathbb E}}}
\nc{\id}{{\mathbb I}}
\nc{\polylog}{\operatorname{polylog}}

\nc{\rnc}{\renewcommand} \nc{\beq}{\begin{equation}}
\nc{\eeq}{{\end{equation}}} \nc{\beqa}{\begin{eqnarray}}
\nc{\eeqa}{\end{eqnarray}} \nc{\lbar}[1]{\overline{#1}}
\nc{\avg}[1]{\langle#1\rangle} \rnc{\max}{\operatorname{max}}
\nc{\rank}{\operatorname{rank}}
\nc{\smfrac}[2]{\mbox{$\frac{#1}{#2}$}} \nc{\ox}{\otimes}
\nc{\dg}{\dagger} \nc{\dn}{\downarrow} \nc{\cA}{{\cal A}}
\nc{\optimal}{^*} \nc{\csupp}{{\operatorname{csupp}}}
\nc{\qsupp}{{\operatorname{qsupp}}}
\nc{\esupp}{{\operatorname{esupp}}} \nc{\var}{\operatorname{var}}
\nc{\rar}{\rightarrow} \nc{\lrar}{\longrightarrow}
\nc{\poly}{\operatorname{poly}}

\usepackage{amsfonts}

\def\01{\{0,1\}}

\newcommand{\eps}{\varepsilon}
\newcommand{\ket}[1]{|#1\rangle}
\newcommand{\bra}[1]{\langle#1|}
\newcommand{\outp}[2]{|#1\rangle\langle#2|}

\newcommand{\braket}[2]{\langle #1|#2\rangle}

\newcommand{\Tr}{\mbox{\rm Tr}}

\newtheorem{definition}{Definition}
\newtheorem{theorem}{Theorem}
\newtheorem{lemma}{Lemma}

\newtheorem{corollary}{Corollary}

\begin{document}

\title{Security of quantum bit string commitment depends on the information measure}

\author{Harry \surname{Buhrman}}
\affiliation{CWI, Kruislaan 413, 1098 SJ
Amsterdam, The Netherlands}

\author{Matthias \surname{Christandl}}
\email[]{matthias.christandl@qubit.org}
\affiliation{Department of Applied Mathematics and Theoretical
Physics, University of Cambridge, Wilberforce Road, Cambridge, CB3
0WA, U.K.}

\author{Patrick \surname{Hayden}}
\affiliation{School of Computer
Science, McGill University, Montreal, Canada}

\author{Hoi-Kwong \surname{Lo}}
\affiliation{Department of
Electrical and Computer Engineering and Department of Physics,
University of Toronto, Toronto, Canada M5G 3G4}

\author{Stephanie \surname{Wehner}}
\email[]{wehner@cwi.nl}
\affiliation{CWI, Kruislaan 413, 1098 SJ
Amsterdam, The Netherlands}

\date{\today}

\begin{abstract}
Unconditionally secure non-relativistic bit commitment is known to
be impossible in both the classical and the quantum world. However,
when committing to a string of $n$ bits at once, how far can we
stretch the quantum limits? In this letter, we introduce a framework
of quantum schemes where Alice commits a string of $n$ bits to Bob,
in such a way that she can only cheat on $a$ bits and Bob can learn
at most $b$ bits of information before the reveal phase. Our results
are two-fold: we show by an explicit construction that in the
traditional approach, where the reveal and guess probabilities form
the security criteria, no good schemes can exist: $a+b$ is at least
$n$. If, however, we use a more liberal criterion of security, the
accessible information, we construct schemes where $a=4\log_2
n+O(1)$ and $b=4$, which is impossible classically. Our findings
significantly extend known no-go results for quantum bit commitment.
\end{abstract}

\maketitle

Imagine two mutually distrustful parties Alice and Bob at distant
locations. They can only communicate over a channel, but want to
play the following game: Alice secretly chooses a bit $x$. Bob wants
to be sure that Alice indeed has made her choice. Yet, Alice wants
to keep $x$ hidden from Bob until she decides to reveal $x$. To
convince Bob that she made up her mind, Alice sends Bob a
commitment. From the commitment alone, Bob cannot deduce $x$. At a
later time, Alice reveals $x$ and enables Bob to open the
commitment. Bob can now check if Alice is telling the truth. This
scenario is known as \emph{bit commitment}. Commitments play a
central role in modern day cryptography. They form an important
building block in the construction of larger protocols in, for
example, gambling and electronic voting, and other instances of secure
two-party computation. In the realm of quantum mechanics, it has
been shown that oblivious transfer~\cite{crepeau:practicalOT} can be
achieved provided there exists a secure bit commitment
scheme~\cite{yao:otFromBc,crepeau:qot}. In turn, classical oblivious
transfer can be used to perform any secure two-party
computation~\cite{crepeau:committedOT}.
Commitments are also useful for constructing
zero-knowledge proofs~\cite{goldreich:book1} and lead to coin
tossing~\cite{blum:coin}.

Classically, unconditionally secure non-relativistic bit commitment
is known to be impossible. Unfortunately, after several quantum
schemes were suggested~\cite{bb84}, non-relativistic {\it quantum}
bit commitment was shown to be impossible,
too~\cite{lo&chau:bitcom,mayers:trouble}, even in the presence of
superselection rules~\cite{kitaev:super}. In fact, only very limited
degrees of concealment and bindingness can be
obtained~\cite{spekkens:tradeoffBc}. It has been shown that the
quantum no-go theorems do not apply to protocols which use two or
more sites and take account of relativistic signaling constraints.
We work in the non-relativistic quantum mechanical setting, hence
all presented results are referring to this setting only. In the
face of the negative results regarding this setting, what can we
still hope to achieve?

In this letter, we consider the task of committing to a string of
$n$ bits at once when both the honest player and the adversary have
unbounded resources. Since perfect bit commitment is impossible,
perfect bit string commitment is impossible, too. We thus give
both Alice and Bob a limited ability to cheat. First, we introduce a
framework for the classification of bit string commitments in terms
of the length $n$ of the string, Alice's ability to cheat on at most
$a$ bits and Bob's ability to acquire at most $b$ bits of
information before the reveal phase. We say that Alice can cheat on
$a$ bits if she can reveal up to $2^a$ strings successfully. Bob's
security definition is crucial to our investigation: If $b$
determines a bound on his probability to guess Alice's string, then
we prove that $a + b$ is at least $n$. This implies that the trivial
protocol, where Alice's commitment consists of sending $b$ bits of
her string to Bob, is optimal. If, however, $b$ is a bound on the
accessible information that the quantum states contain about Alice's
string, then we show that non-trivial schemes exist. More precisely,
we construct schemes with $a=4\log_2 n+O(1)$ and $b=4$. This is
impossible classically.

Quantum commitments of strings have previously been considered by
Kent~\cite{kent:sc}, who pointed out that in the quantum world
useful bit string commitments could be possible despite the no-go
theorem for bit commitment. His scenario differs significantly from
ours and imposes an additional constraint, which is not present in
our work: Alice does not commit to a superposition of strings.
\\

\noindent {\it Framework.}
\begin{definition} \emph{An $(n,a,b)$-\emph{Quantum Bit String Commitment (QBSC)} is a
quantum communication protocol between two parties, Alice (the
committer) and Bob (the receiver), which consists of two phases and
two security requirements.
\begin{itemize}
\item (Commit Phase) Assume that both parties are honest.
Alice chooses a string $x \in \01^n$ with probability $p_x$.
Alice and Bob communicate and at the end Bob holds state $\rho_x$.
\item (Reveal Phase)
If both parties are honest, Alice and Bob communicate and at the end
Bob learns $x$. Bob accepts.
\item (Concealing)
If Alice is honest, $\sum_{x \in \01^n}p^B_{x|x} \leq 2^{b}$, where
$p^B_{x|x}$ is the probability that Bob correctly guesses $x$ before
the reveal phase given $\rho_x$.
\item (Binding)
If Bob is honest, then for all commitments of Alice: $\sum_{x \in
\01^n} p^A_x \leq 2^a$, where $p^A_x$ is the probability that Alice
successfully reveals $x$.
\end{itemize}}
\end{definition}
We say that Alice \emph{successfully reveals} a string $x$ if Bob accepts
the opening of $x$, i.e. he performs a test depending on the individual
protocol to check Alice's honesty and concludes that she was indeed honest.
Note that quantumly, Alice can always
commit to a superposition of different strings without being
detected. Thus even for a perfectly binding bit string commitment
(i.e. $a=0$) we only demand that $\sum_{x \in \{0,1\}^n}p^A_x \leq 1$,
whereas classically one wants that $p^A_{x'}=\delta_{x, x'}$.
Note that our concealing definition reflects Bob's a priori
knowledge about $x$. We choose an a priori uniform distribution
(i.e. $p_x = 2^{-n}$) for $(n,a,b)$-QBSCs, which naturally comes from the
fact that we consider $n$-bit strings. A generalization to any
$(P_X,a,b)$-QBSC where $P_X$ is an arbitrary distribution is
possible but omitted in order not to obscure our main line of
argument.
Instead of Bob's guessing probability, one can take any information
measure $B$ to express the security against Bob. In general, we consider an
$(n,a,b)$-$\mbox{QBSC}_B$ where the new Concealing-condition reads
$B(\ens) \leq b$ with ensemble $\ens = \{p_x,\rho_x\}$. In the
latter part of this letter we show that for $B$ being the
\emph{accessible information} non-trivial protocols, i.e. protocols
with $a+b \ll n$, exist. The accessible information is defined as
$I_{acc}(\ens) = \max_{M} I(X;Y)$, where $P_X$ is the prior
distribution of the random variable $X$, $Y$ is the random variable
of the outcome of Bob's measurement on $\ens$, and the
maximization is taken over all measurements $M$.
\\

\noindent {\it Impossibility.} Our impossibility result will be
proven in three steps: we first show that any $(n,a,b)$-QBSC is also
an $(n,a,b)$-$\mbox{QBSC}_\xi$ with the security measure $\xi(\ens)$
defined below in eq.~(\ref{xi}). Secondly, we prove that an
$(n,a,b)$-$\mbox{QBSC}_\xi$ can only exist for values $a$, $b$ and
$n$ obeying $a + b + c \geq n$, where $c$ is a small constant
independent of $a$, $b$ and $n$. This in turn implies the
impossibility of an $(n,a,b)$-QBSC for such parameters. Finally, we
show that $\emph{many}$ executions of the protocol can only be
secure if $a + b \geq n$. Before we proceed to the proof, we
introduce a few tools from quantum information theory.

We work in the model of two-party non-relativistic quantum protocols of
Yao~\cite{yao:otFromBc}, simplified by Lo and
Chau~\cite{lo&chau:bitcom} which is usually adopted in this context.
Here, any two-party quantum protocol can be regarded as a pair of
quantum machines (Alice and Bob), interacting through a quantum channel.
Consider the product of
three Hilbert spaces $\hilbert_A$, $\hilbert_B$ and $\hilbert_C$ of
bounded dimensions representing the Hilbert spaces of Alice's and
Bob's machines and the channel, respectively. Without loss of
generality, we assume that each machine is initially in a specified
pure state. Alice and Bob perform a number of rounds of
communication over the channel. Each such round can be modeled as a
unitary transformation on $\mathcal{H}_A \otimes \mathcal{H}_C$ and
$\mathcal{H}_B \otimes \mathcal{H}_C$ respectively.
Since the protocol is known to both Alice and Bob, they know
the set of possible unitary transformations used in the protocol.
We assume that Alice and Bob are in
possession of both a quantum computer and a quantum storage device.
This enables them to add ancillae to the quantum machine and use
reversible unitary operations to replace measurements. By doing so,
Alice and Bob can delay measurements and thus we can limit ourselves
to protocols where both parties only measure at the
very end. Moreover, any classical computation or communication that
may occur can be simulated by a quantum computer.

We now show that every $(n,a,b)$-$\mbox{QBSC}$ is an
$(n,a,b)$-$\mbox{QBSC}_\xi$. The security measure $\xi(\ens)$ is
defined by
\begin{equation} \label{xi}
\xi (\ens)\equiv n- H_2(\rho_{AB}|\rho),
\end{equation}
where $\rho_{AB} = \sum_x p_x \outp{x}{x} \otimes \rho_x$ and $\rho
= \sum_x p_x \rho_x$ are only dependent on the ensemble $\ens=\{p_x,
\rho_x\}$. $H_2(\cdot|\cdot)$ is an entropic quantity defined
in~\cite{renato:diss} $H_2(\rho^{AB}|\rho)\equiv -\log\Tr ((\id
\otimes \rho^{-\frac{1}{2}})\rho_{AB})^2.$ This quantity is directly
connected to Bob's maximal average probability of successful guessing
the string:
\begin{lemma} \label{guessing-lemma} Bob's maximal average probability of successfully guessing
the committed string,~i.e. $\sup_M \sum_x p_x p^{B, M}_{x|x}$ where $M$
ranges over all measurements and $p^{B, M}_{y|x}$ is the conditional
probability of guessing $y$ given $\rho_x$, is larger or equal to
$2^{- H_2(\rho_{AB}|\rho)}$.
\end{lemma} {\it Proof: } By definition the maximum average guessing
probability is lower bounded by the average guessing probability for
a particular measurement strategy. We choose the \emph{square-root
measurement} which has operators $M_x = p_x \rho^{-\frac{1}{2}}
\rho_x \rho^{-\frac{1}{2}}$. $p^B_{x|x} = \Tr(M_x \rho_x)$ is the
probability that Bob guesses $x$ given $\rho_x$, hence
\begin{eqnarray*}
\log_2\sum_x p_x p^{B,\max}_{x|x}
&\geq& \log_2\sum_x p_x^2 \Tr(\rho^{-\frac{1}{2}} \rho_x \rho^{-\frac{1}{2}} \rho_x)\\
&=& \log_2\Tr \left(\left[(\id \otimes \rho^{-\frac{1}{2}})\rho_{AB}\right]^2\right)\\
&=& - H_2(\rho_{AB}|\rho).\hspace{3.5cm} \Box
\end{eqnarray*}
Related estimates were derived in~\cite{barnumknill}. For the uniform
distribution $p_x = 2^{-n}$ we have from the concealing
condition that $\sum_x p^B_{x|x} \leq 2^{b}$ which by
Lemma~\ref{guessing-lemma} implies $\xi(\ens) \leq b$. Thus, every
$(n,a,b)$-QBSC is an $(n,a,b)$-$\mbox{QBSC}_\xi$.

We make use of the following theorem, known as \emph{privacy
amplification against a quantum adversary}. In our case, Bob holds
the quantum memory and privacy amplification is used to find Alice's
attack.
\begin{theorem}[Th.~5.5.1 in~\cite{renato:diss} (see
also~\cite{KoMaRe05})]\label{theorem:renato} Let $\mathcal{G}$ be a
class of two-universal hash functions~\footnote{For instance, the
class of all functions from $\01^n$ to $\01^s$ is
\emph{two-universal}. For a definition and efficient constructions
see~\cite{CarWeg79}.} from $\01^n$ to $\01^s$. Application of $g \in
\mathcal{G}$ to the random variable $X$ maps the ensemble
$\ens=\{p_x, \rho_x\}$ to $\cE_g=\{q^g_y, \sigma^g_y\}$ with
probabilities $q^g_y = \sum_{x \in g^{-1}(y)} p_x$ and quantum
states $\sigma^g_y = \sum_{x \in g^{-1}(y)} p_x \rho_x$. Then \be
\label{eq-renner-koenig} \frac{1}{|\mathcal{G}|} \sum_{g \in
\mathcal{G}} d(\cE_g)  \leq
\frac{1}{2}2^{-\frac{1}{2}[H_2(\rho_{AB}|\rho) - s]},\ee where
$d(\cE) \equiv \delta\big(\sum_x p_x \proj{x} \otimes \rho_x,
\id/2^n\otimes \rho\big)$ (and similarly for $d(\cE_g)$) and
$\delta(\alpha, \beta)\equiv\frac{1}{2} \Tr |\alpha-\beta|$.
\end{theorem}

\par
The following reasoning is used to prove the impossibility of
quantum bit commitment~\cite{lo&chau:bitcom, mayers:trouble}:
Suppose $\rho_0$ and $\rho_1$ are density operators that correspond
to a commitment of a ``0'' or a ``1'' respectively. Let
$\ket{\phi_0}$ and $\ket{\phi_1}$ be the corresponding purifications
on the joint system of Alice and Bob. If $\rho_0$ equals $\rho_1$
then Alice can find a local unitary transformation $U$ that Alice
can apply to her part of the system such that
$\ket{\phi_1} = U \otimes \id \ket{\phi_0}$. This enables Alice to change the total
state from $\ket{\phi_0}$ to $\ket{\phi_1}$ and thus cheat. This
reasoning also holds in an approximate
sense~\cite{mayers:trouble}, here used in the following
form~\footnote{\emph{Proof of Lemma~\ref{lemma:mayers}:}
$\delta(\rho_0, \rho_1) \leq \epsilon$ implies $F(\rho_0, \rho_1)
\geq 1-\epsilon$. $F(\cdot, \cdot)$ is the fidelity of two quantum
states, which, by Uhlmann's theorem equals $\max_{U}| \bra{\phi_0} U
\otimes \id\ket{\phi_1}|$. Here, $\ket{\phi_0}$ and $\ket{\phi_1}$
are the joint states after the commit phase and the maximization
ranges over all unitaries $U$ on Alice's (i.e. the purification)
side. Let $\ket{\psi_0} = U \otimes \id \ket{\phi_1}$ for a $U$
achieving the maximization. Then $\delta(\proj{\phi_0},
\proj{\psi_0})=\sqrt{1-|\braket{\phi_0}{\psi_0}|}\leq
\sqrt{1-(1-\epsilon)^2}\leq \sqrt{2\epsilon}$. If both parties are
honest, the reveal phase can be regarded as a measurement resulting
in a distribution $P_Y$ ($P_Z$) if $\ket{\phi_0}$ ($\ket{\psi_0}$)
was the state before the reveal phase. The random variables $Y$ and
$Z$ carry the opened bit or the value `reject (r)'. Since the trace
distance does not increase under measurements, $\delta(P_Y, P_Z)
\leq \delta(\proj{\phi_0}, \proj{\psi_0}) \leq \sqrt{2\epsilon}$.
Hence $\frac{1}{2}(|P_Y(0)-P_Z(0)| +
|P_Y(1)-P_Z(1)|+|P_Y(r)-P_Z(r)|)  \leq \sqrt{2\epsilon}$. Since
$\ket{\phi_0}$ corresponds to Alice's honest commitment to $0$ we
have $P_Y(0)=1$, $P_Y(1)=P_Y(r)=0$ and hence $P_Z(0)\geq
1-\sqrt{2\epsilon}$. \hfill $\Box$ }:
\begin{lemma}\label{lemma:mayers}
Let $\delta(\rho_0, \rho_1)\leq \epsilon$ and assume that the
bit-commitment protocol is error-free if both parties are honest. Then there is a
method for Alice to cheat such that the probability of successfully
revealing a $0$ given that she committed to a $1$ is greater or
equal to $1-\sqrt{2\epsilon}$.
\end{lemma}
Now, we can prove our impossibility result.

\begin{theorem} \label{nogo}$(n,a,b)$-$\mbox{QBSC}_\xi$ schemes, and thus
also $(n,a,b)$-$\mbox{QBSC}$ schemes, with $a+b+c < n$ do not exist.
$c$ is a constant equal to $5 \log_2 5 - 4\approx 7.61$.
\end{theorem}
{\it Proof:} Consider an $(n,a,b)$-$\mbox{QBSC}_\xi$ and the case
where both Alice and Bob are honest. Alice committed to
$x$. We denote the joint state of the system Alice-Bob-Channel
$\hilbert_A \otimes \hilbert_B \otimes \hilbert_C$ after the commit
phase by $\ket{\phi_x}$ for input state $\ket{x}$.
Let $\rho_x$ be Bob's reduced density matrix, and let
$\cE=\{p_x, \rho_x\}$
where $p_x=2^{-n}$.

Assuming that Bob is honest, we will give a cheating strategy for
Alice in the case where $a+b+5 \log_2 5 - 4 < n$. The strategy will
depend on the two-universal hash function $g:\mathcal{X} = \01^n \rightarrow
\mathcal{Y} = \01^{n -m}$, for appropriately chosen $m$. Alice
picks a $y\in \mathcal{Y}$ and constructs the state $ (\sum_{x \in
g^{-1}(y)}\ket{x}\ket{x})/\sqrt{|g^{-1}(y)|}$. She then gives the
second half of this state as input to the protocol and stays honest
for the rest of the commit phase. The joint state of Alice and Bob
at the end of the commit phase is thus $\ket{\psi^g_y} =
(\sum_{x \in g^{-1}(y)}\ket{x}\ket{\phi_x})/\sqrt{|g^{-1}(y)|}$. The
reduced states on Bob's side are
$\sigma^g_y = \frac{1}{q^g_y}\sum_{x \in g^{-1}(y)} p_x \rho_x$
with probability $q^g_y = \sum_{x \in g^{-1}(y)} p_x$. We
denote this ensemble by $\cE_g$. Let $\sigma =
\sigma^g = \sum_y q^g_y \sigma^g_y$ for all $g$.

We now apply Theorem~\ref{theorem:renato} with $s=n-m$ and
$\xi(\ens)\leq b$ and obtain $ \frac{1}{|\mathcal{G}|} \sum_{g \in
\mathcal{G}} d(\cE_g) \leq \eps $
where $\eps=\frac{1}{2}2^{-\frac{1}{2}(m-b)}$. Hence, there is at
least one $g$ such that $d(\cE_g) \leq \eps$; intuitively, this
means that Bob knows only very little about the value of $g(x)$.
This $g$ defines Alice's cheating strategy. It is straightforward to verify that
$d(\cE_g) \leq \eps$ implies
\begin{equation}\label{sum_eq}
2^{-(n-m)} \sum_{y} \delta(\sigma, \sigma^g_y) \leq 2\eps.
\end{equation}
Let us therefore assume without loss of generality that Alice
chooses $y_0\in \cY$ with $\delta(\sigma^g_{y_0}, \sigma) \leq
2\eps$.

Clearly, the probability to successfully reveal some $x$ in
$g^{-1}(y)$ given $\ket{\psi^g_y}$ is one~\footnote{Alice learns
$x$, but cannot pick it: she committed to a superposition and $x$ is
chosen randomly by the measurement.}. Thus the probability to reveal
$y$ (i.e. to reveal an $x$ such that $y = g(x)$) given $\ket{\psi^g_y}$
successfully is one. Let $\tilde{p}_x$
and $\tilde{q}^g_y$ denote the probabilities to successfully reveal
$x$ and $y$ respectively and $\tilde{p}^g_{x|y}$ be the conditional
probability to successfully reveal $x$, given $y$. We have
$\sum_x \tilde{p}_x =\sum_y \tilde{q}^g_y
\sum_{x \in g^{-1}(y)}\tilde{p}^g_{x|y} \geq \sum_y \tilde{q}^g_y$.

Recall that Alice can transform $\ket{\psi^g_{y_0}}$ approximately
into $\ket{\psi^g_y}$ if $\sigma^g_{y_0}$ is sufficiently close to
$\sigma^g_y$ by using only local transformations on her part. It
follows from Lemma~\ref{lemma:mayers} that we can estimate the
probability of revealing $y$, given that the state was really
$\ket{\psi_{y_0}}$. Since this reasoning applies to all $y$, on
average, we have
\begin{eqnarray*}
\sum_y \tilde{q}^g_y &\geq& \sum_y (1- 2^{\frac{1}{2}} \delta(\sigma^g_{y_0}, \sigma^g_{y})^{\frac{1}{2}})  \\
&\geq& 2^{n-m} - 2^{\frac{1}{2}} 2^{n-m} \big(2^{m-n} \sum_y \delta(\sigma^g_{y_0},\sigma^g_{y})\big)^{\frac{1}{2}} \\
&\geq& 2^{n-m}[1 - 2^{\frac{1}{2}} (2^{m-n}
(\sum_y \delta(\sigma^g_{y_0},\sigma) + \delta(\sigma,\sigma^g_{y})))^{\frac{1}{2}}]\\
&\geq& 2^{n-m}(1 - 2 (2\eps)^{\frac{1}{2}}),
\end{eqnarray*} where the first inequality follows from
Lemma~\ref{lemma:mayers}, the second from Jensen's inequality and
the concavity of the square root function, the third from the
triangle inequality and the fourth from eq.~(\ref{sum_eq}) and
$\delta(\sigma^g_{y_0}, \sigma) \leq 2\eps$. Recall that to be
secure against Alice, we require $2^a \geq 2^{n-m}(1- 2
(2\eps)^{\frac{1}{2}})$. We insert
$\epsilon=\frac{1}{2}2^{-\frac{1}{2}(m-b)}$, define $m=b+\gamma$ and
take the logarithm on both sides to get \be \label{eq-delta} a+b
+\delta \geq n,\ee where $\delta = \gamma -\log (1-2^{-\gamma/4
+1})$. Keeping in mind that $1-2^{-\gamma/4 +1}>0$ (or equivalently
$\gamma>4$), we find that the minimum value of $\delta$ for which
eq.~(\ref{eq-delta}) is satisfied is $\delta=5 \log_2 5 -4$ and
arises from $\gamma = 4 (\log_2 5 - 1)$. Thus,
no $(n,a,b)$-$\mbox{QBSC}_\xi$ with $a+b+5 \log_2 5 - 4 < n$ exists.
\hfill $\Box$
\par
Since the constant $c$ does not depend on $a$, $b$ and $n$, multiple
parallel executions of the protocol can only be secure if $a+b\geq
n$:
\begin{corollary} Let $P$ be an $(n,a,b)$-QBSC with $P^m$ an
$(mn,ma,mb)$-QBSC. Then $n < a + b + c/m$. In particular, no $(n, a,
b)$-QBSC with $a+b<n$ can be executed securely an arbitrary number
of times in parallel. The latter statement also applies to $(n, a,
b)-\mbox{QBSC}_{\chi}$s, where $\chi$ denotes the Holevo information
of the ensemble $\cE$~\cite{BCHLW05}.
\end{corollary} It follows directly from~\cite{kitaev:super} that
the results in this section also hold in the presence of
superselection rules.
\\

\noindent {\it Possibility.} Surprisingly, if one is willing to
measure Bob's ability to learn $x$ using the accessible information
non-trivial protocols become possible. These protocols are based on
a discovery known as ``locking of classical information in quantum
states''~\cite{terhal:locking}.
The protocol, which we call LOCKCOM($n$, $\cU$), uses this effect
and is specified by a set $\cU = \{U_1,\ldots,U_{|\cU|}\}$ of
unitaries.
\begin{itemize}
\item Commit phase: Alice has the string $x \in \01^n$ and randomly chooses
$r \in \{1,\ldots,|\cU|\}$. She sends the state $U_r \ket{x}$ to
Bob, where $U_r \in \cU$.
\item Reveal phase: Alice announces $r$ and $x$. Bob applies
$U_r^\dagger$ and measures in the computational basis
to obtain $x'$. He accepts if and only if $x'=x$.
\end{itemize}
As a first observation, the number of unitaries $|\cU|$ limits the
number of different ways of revealing a string, i.e. $2^a \leq
|\cU|$~\footnote{This can be seen as follows. Let $\tilde{p}_x$
denote the probability that Alice reveals $x$ successfully. Then,
$\tilde{p}_x \leq \sum_r \tilde{p}_{x,r}$, where $\tilde{p}_{x,r}$
is the probability that $x$ is accepted by Bob when the reveal
information was $r$. Let $\rho$ denote the state of Bob's system.
Summation over $x$ yields $
\sum_x \tilde{p}_x \leq  \sum_{x, r}\tilde{p}_{x,r} = \sum_{x,r} \Tr \proj{x} U_r^\dagger \rho U_r = \sum_r \Tr \rho = 2^a.\\
$}. Furthermore we have adapted the work
in~\cite{winter:randomizing} in order to show that there exist
$O(n^4)$ unitaries that bring Bob's accessible information down to a
constant: $\I(\cE) \leq 4$~\cite[Appendix
B.2]{BCHLW05}. In summary:
\begin{theorem} \label{theorem-LOCKCOM}
For $n \geq 3$, there exist $(n, 4\log_2 n +O(1), 4)$-$
\mbox{QBSC}_{\I}$ protocols.
\end{theorem}
The protocol is as follows:
Alice chooses a set of $O(n^4)$ unitaries independently
according to the Haar measure (approximated) and announces the resulting set $\cU$
to Bob. They then perform LOCKCOM($n,\cU$). Our analysis shows that
this variant is secure against Bob with high probability.
Unfortunately, the protocol is inefficient both in terms of
computation and communication. It remains open to find an efficient
constructive scheme with those parameters.

In contrast, for only two bases, an efficient construction exists
and uses the identity and the Hadamard transform as unitaries.
From~\cite{terhal:locking} (see also~\cite{CW05}) it then follows
that LOCKCOM($n, \{\id^{\otimes n}, H^{\otimes n}\})$ is an $(n, 1,
n/2)$-$\mbox{QBSC}_{\I}$ protocol. As shown
in~\cite{matthias:thesis}, this protocol can be made \emph{cheat
sensitive}~\cite{yao:bitEscrow} for Bob, i.e. \emph{any} nonzero
information-gain by Bob will be detected by Alice with nonzero
probability.

A drawback of weakening the security requirement is that LOCKCOM
protocols are not necessarily composable. Therefore, if LOCKCOM is
used as a sub-protocol in a larger protocol, the security of the
resulting scheme has to be evaluated on a case by case basis.
However, LOCKCOM protocols are secure when
executed in parallel. This is a consequence of the definition of
Alice's security parameter and the additivity of the accessible
information~\cite{DLT02}, and sufficient for many cryptographic
purposes.
\\
\noindent {\it Conclusion} We have introduced a framework for
quantum commitments to a string of bits and shown that under strong
security requirements (e.g. bounded guessing probability or Holevo
information), non-trivial protocols do not exist. A property of
quantum states known as \emph{locking}, however, allowed us to
propose meaningful protocols for a weaker security demand: Alice
encodes her classical $n$ bit string into a quantum state in such a
way that no measurement on Bob's side will yield high mutual
information with the commitment. Alice is genuinely committed,
because the quantum states that she sent contain almost the
complete commitment, i.e. have high Holevo information.

\noindent {\it Acknowledgments} We thank J.~Barrett, I.~Damg{\aa}rd,
A.~Kent, S.~Massar, R.~Renner and R.~Spekkens for discussions and
R.~de Wolf, A.~Broadbent and an anonymous referee for helpful
comments. We also thank R.~Jain for discussion on his
work~\cite{Jain05}, where, following our preprint~\cite{BCHLW05}, he
used a different method to prove that $(n,a,b)$-$\mbox{QBSC}_\chi$s
with $a + 16b + 31 < n$, do not exist. The attack that he constructs
for Alice in order to prove his result, however, aborts with high
probably. In the proof of Theorem~\ref{nogo}, in contrast, Alice's
cheating is only detected with negligible probability. C.~Mochon has
pointed out to us that Kitaev's lower bound for coin flipping can be
used to achieve similar no-go result than we have presented in this
letter. However, our no-go leads to an explicit attack by Alice.

M.C. was supported by a DAAD Doktorandenstipen\-dium, the EPSRC and
a Magdalene College Nevile Research Fellowship. P.H. and H.-K.L. are
supported by the Canadian funding agencies CFI, CIAR, CIPI, CRC,
NSERC, PREA and OIT. H.B. and S.W. are supported by the NWO vici
project 2004-2009. We acknowledge support from EU project RESQ
IST-2001-37559, QAP IST 015848 and the FP6-FET Integrated Project SCALA, CT-015714.

\end{document}